# A Model for Grain Boundary Thermodynamics

Reza Darvishi Kamachali

Federal Institute for Materials Research and Testing (BAM), Unter den Eichen 87, 12205 Berlin
Max-Planck-Institut für Eisenforschung (MPIE), Max-Planck-Str. 1, 40237 Düsseldorf
reza.kamachali@bam.de; reza.kamachali@gmail.com

Systematic microstructure design requires reliable thermodynamic descriptions of each and all microstructure elements. While such descriptions are well established for most bulk phases, thermodynamic assessment of crystal defects is challenged because of their individualistic nature. In this paper, a model is devised for assessing grain boundary thermodynamics based on available bulk thermodynamic data. We propose a continuous relative atomic density field and its spatial gradients to describe the grain boundary region with reference to the defect-free, homogeneous bulk and derive the grain boundary Gibbs free energy functional. Grain boundary segregation isotherm and phase diagram are computed for a regular binary solid solution, qualitatively discussed for the Pt-Au system. The relationships between the grain boundary's atomic density, excess free volume, and misorientation angle are discussed. Combining the current density-based model with available bulk thermodynamic databases capacitates constructing databases, phase diagrams, and segregation isotherms for grain boundaries, opening possibilities for studying and designing heterogeneous microstructures.

# 1. Introduction

Grain boundaries (GBs) are one of the main sources of heterogeneity in polycrystalline microstructures; They interact with various microstructure elements, e.g. GBs' network [1], secondary-phase particles [2-4], dislocations [5-7], and vacancies [8, 9], while they themselves possess distinct physical properties similar to the bulk phases. GBs can also interact with solute atoms that influences both local and global chemistry of the microstructure. Solute segregation to GBs can mediate a whole different sort of phenomena such as segregation—assisted GB premelting [10], phase transition [11-13], precipitation [14, 15] and embrittlement [16, 17] as well as stabilization of nanocrystalline materials [18-20]. Hence, understanding and controlling GBs and their spatial interactions with other microstructure elements can unroll new ideas for designing heterogeneous microstructures.

The studies on interfacial segregation can be traced back to the works of Gibbs [21], Langmuir [22], McLean [23] and Fowler and Guggenheim [24] on the adsorption. Since then, numerous studies on the segregation phenomena have been conducted [25-42]. Van der Waals [43], Cahn and Hilliard [44] and Cahn [45] worked out free energy functionals that account for heterogeneous interfacial features on the mesoscale. Based on this idea, several models for GB segregation were proposed. Ma *et al* [46] developed a model for studying GB segregation in which GB region is described by its reduced atomic coordination number. Based on Kobayashi-Warren-Carter's work [47], Tang *et al* developed models for order-disorder transition [48, 49] and phase transition [50] at GBs. Heo *et al* [51] proposed a model for GB segregation and drag including misfit elastic interactions. The common feature of these models is the non-vanishing gradients in GB structure and/or composition. Due to this fact, GB phases are sometimes referred to as 'complexions' [52-54] to emphasize their confined heterogeneous nature in contrast to a bulk phase that is –by definition– homogeneous.

Despite the enormous knowledge accumulated over the last century about the fundamental aspects of GBs [39, 40, 55, 56], its application at a technical level is awaiting more comprehensive tools to allow systematic GB engineering. In particular, databases on GBs' properties are less popular; While thermodynamic and kinetic descriptions for most bulk phases are well established, analogous systematic and general descriptions for GBs are rarely investigated. This is because of the individualistic nature of GBs as every GB can be different depending on its crystallographic properties [55]. Although GB properties are functions of a large space of crystallographic variables, one may not neglect the fact that they are made of the same constituents as for the grain interior (bulk), albeit deviating in their structure. An alternative view, therefore, can be to picture GBs with reference to their corresponding bulk phases. Thus, parallel to the current characterization techniques [57] and automated simulation methods [58, 59] that are mainly focused on studying individual GBs, a general approach could be assessing the thermodynamic and kinetic properties of the GBs with reference to the known bulk. For this purpose, we need to establish a physical framework that allows an approximation of the GB environment with respect to its reference bulk material.

In the current study, I propose a continuous atomic density field and its spatial variations to describe the GB region, relative to its adjacent homogeneous bulk material, in a regular substitutional binary solution. Based on this idea, (i) the GB's Gibbs free energy is approximated, (ii) GB segregation and the coexistence of the bulk and GB phases are discussed and, (iii) a concept for computing GB phase diagram is described. The novelty of the current model is that it enables a rapid, pragmatic assessment of GB properties based on available bulk thermodynamic data. To demonstrate the application of the current model, GB segregation and interfacial phase separation in the Pt-Au system are qualitatively studied. Furthermore, the relationships between the relative atomic density field and the GB nature in terms of GB excess free volume and misorientation angle are discussed. In two parallel studies, the applications of this model are demonstrated in understanding segregation and interfacial phase separation in different alloy systems. In one case, we apply the density-based formulation to realize the Fowler-Guggenheim segregation isotherm for co-segregating Ni and Mn atoms at various types of GBs in a FeMnNiCrCo high-entropy alloy [60]. A simpler version of the model has been applied to studying Mn segregation in the binary FeMn alloy system [61]. In this alloy system, we have revealed a *transient* interfacial spinodal phase separation with potential implications to segregation engineering through desired heat treatment processes. The current density-based model offers a pragmatic approach to build GB thermodynamic databases, segregation isotherms and phase diagrams for studying heterogeneous polycrystalline materials.

## 2. Density-based Model for GB Thermodynamics

In his seminal work, van der Waals proposed that an interface can be described by a continuous density field [43] and its spatial variations. For a GB, that needs to accommodate for the geometrical mismatch between two adjacent grains, an analogous picture can be portrayed in which the density of the GB will be different than that of the bulk materials. This is, on the one hand, even a simpler case than the van der Waals's setup, as the bulk density on the two sides of the GB are the same. On the other hand, however, it can be more complex because of the crystalline nature of the bulk material that may extend into the GB region. As a first attempt, here we neglect the heterogeneity of GB density (the density variation within the GB plane due to the crystallinity) and instead focus on the possibility of considering an 'average' atomic density parameter and its gradient terms to describe GB region.

In the following, we derive the density-based Gibbs free energy of a GB in unary (Sec. 2.1) and a substitutional binary regular solution (Sec. 2.2) systems. To do so, we take a variational approach to obtain density-based free energy density and cast this into the conventional thermodynamic framework as used in CALPHAD approaches. We obtain the Gibbs free energy functional as a function of the atomic density field, concentration field, and their respective spatial variations. Using this free energy description, we obtain the GB phase diagram and segregation isotherm, Sec. 3. The model is demonstrated on the Pt-Au system. Different aspects of the current model, i.e., the coexistence of the bulk and GB phases and the relationship between the GB density, GB excess free volume, and GB misorientation angle, are discussed in Sec. 4.

**2.1 A GB in a pure substance**

At constant temperature $T$ and pressure $p$, the Gibbs free energy functional of a system made of pure substance A can be written as

$$\mathcal{G}_A = \int_\Omega G_A \, d\vec{r} = \int_\Omega \rho_n \left( H_A - TS_A \right) d\vec{r} \tag{1}$$

where $\rho_n(\vec{r})$ is the atomic density field equivalent to the inverse molar volume field $V_m^{-1}(\vec{r})$, $H_A$ is enthalpy per unit mole with $H_A = K_A + E_A + pV_A$ ($K_A$: kinetic energy, $E_A$: potential energy and $pV_A$: mechanical energy), and $S_A$ is entropy per unit mole, respectively. Van der Waals has shown [43] that the potential energy of a heterogeneous system depends on the (local) density as well as (nonlocal) density gradients. In this study, we consider a symmetric flat GB separating two infinitely large homogeneous grains. For this system, breaking the symmetry only normal to the GB plane, the interactions can be realized considering only one spatial dimension. The potential energy density will be

$$E_A(x) = E_A(-\infty) + \frac{1}{2} \int_{-\infty}^{x} f(r) \, dr. \tag{2}$$

Here $E_A(-\infty)$ is the potential energy inside the homogeneous grain at $x = -\infty$ and the second term describes the energy stored in matter when brought from $-\infty$ to a given position $x$. $f(r)$ is the sum of all forces acting on point $r$:

$$f(r) = \int_0^\infty [\zeta(r+q) - \zeta(r-q)] \, dq \tag{3}$$

where $\zeta(r \pm q)$ is the force density at point $r$ upon the interaction between the two material points separated by a distance $\pm q$. The interaction forces between atoms depend on the interatomic potential $U(r \pm q)$ with $\zeta(r \pm q) = \frac{\partial U(r \pm q)}{\partial q}$.

Although the detailed form of the interatomic potentials/forces can vary for different types of atoms, it is well-known that the atomistic interactions rapidly decays over the distance between atoms. For the sake of our discussion, we consider here a simple functional form

$$\zeta(r \pm q) = \frac{\alpha\, n(r \pm q)}{q^z} = \psi(q)\, \rho_n(r \pm q) \qquad (4)$$

where $\alpha$ is a material constant, $n(r \pm q)$ is the number of atoms, $z$ is a positive exponent and we define $\psi(q) = \frac{\alpha V_L}{q^z}$. Equation (4) with $z \gg 1$ is inspired by interatomic force relations in the absence of the long-range electrostatic interactions [62]. Here $V_L$ is the characteristic coarse-graining volume over which the mesoscale atomic density field is measured by $\rho_n(r) = \frac{n(r)}{V_L}$. Similar coarse-graining descriptions are worked out in defining mesoscale free energy formulations [63, 64]. The specific form of $\psi(q)$ can be obtained from atomistic simulations. The second- or higher-order dependencies of the interaction forces on the atomic density (including repulsive atomic interactions) can also be assumed. In the current treatment, however, we limit ourselves to the simplest first-order linear approximation as in Eq. (4).

In the theories of atomistic simulations, the atomic forces are practically calculated up to a cut-off radius $R_c$ beyond which the interatomic forces are neglected [62]. Hence, one reasonable approximation for the coarse-graining volume would be $V_L \approx \frac{4\pi}{3} R_c^3$ which can relate the atomistic simulation length-scale with the current mesoscale density-based formulation. Inserting Eq. (4) in Eq. (3), and integrating by part we can write

$$\begin{aligned}
f(r) &= \int_0^\infty \psi(q)\, [\rho_n(r+q) - \rho_n(r-q)]\, \mathrm{d}q \\
&= \psi(q)\, [\rho_n(r+q) - \rho_n(r-q)]\Big|_0^\infty - \int_0^\infty \psi(q)\, \frac{\partial[\rho_n(r+q) - \rho_n(r-q)]}{\partial q}\, \mathrm{d}q \\
&= -\int_0^\infty \psi(q)\, \frac{\partial[\rho_n(r+q) - \rho_n(r-q)]}{\partial q}\, \mathrm{d}q
\end{aligned} \qquad (5)$$

with $\lim_{q \to \infty} \psi(q) = 0$ and $\lim_{q \to 0} [\rho_n(r+q) - \rho_n(r-q)] = 0$. Applying the Taylor expansions $\rho_n(r \pm q) = \rho_n(r) \pm \frac{\partial \rho_n}{\partial r} q + \frac{1}{2!}\frac{\partial^2 \rho_n}{\partial r^2} q^2 \pm \frac{1}{3!}\frac{\partial^3 \rho_n}{\partial r^3} q^3 + \frac{1}{4!}\frac{\partial^4 \rho_n}{\partial r^4} q^4 \pm \frac{1}{5!}\frac{\partial^5 \rho_n}{\partial r^5} q^5 + \cdots$, Eq. (5) gives

$$f(r) = -\int_0^\infty 2\,\psi(q)\left[\frac{\partial \rho_n}{\partial r} + \frac{1}{2!}\frac{\partial^3 \rho_n}{\partial r^3}q^2 + \frac{1}{4!}\frac{\partial^5 \rho_n}{\partial r^5}q^4 + \cdots\right] dq$$
$$= 2E_A^0\left[\frac{\partial \rho_n}{\partial r}\right] - \kappa_A\left[\frac{\partial^3 \rho_n}{\partial r^3}\right] - \kappa_A'\left[\frac{\partial^5 \rho_n}{\partial r^5}\right] - \cdots \quad (6)$$

where we define $E_A^0 = -\int_0^\infty \psi(q)\,dq$, $\kappa_A = \int_0^\infty \psi(q)q^2\,dq$ and $\kappa_A' = \frac{1}{12}\int_0^\infty \psi(q)q^4\,dq$. Using Eqs. (2) and (6) with the boundary values $E_A(-\infty) = E_A^0 \rho_n(-\infty)$, $\left(\frac{\partial^2 \rho_n}{\partial r^2}\right)_{-\infty} = 0$ and $\left(\frac{\partial^4 \rho_n}{\partial r^4}\right)_{-\infty} = 0$, we obtain the potential energy as a function of atomic density:

$$E_A(x) = E_A^0 \rho_n(x) - \frac{\kappa_A}{2}\left(\frac{\partial^2 \rho_n}{\partial r^2}\right)_x - \frac{\kappa_A'}{2}\left(\frac{\partial^4 \rho_n}{\partial r^4}\right)_x - \cdots \quad (7)$$

As a consequence of the Tylor expansions, it is clear that the sixth and higher-order spatial derivatives of the atomic density field can still contribute to the potential energy. These are, however, neglected in the following. Thus, the free energy functional in Eq. (1) can be written as

$$\mathcal{G}_A = \int_\Omega \left[E_A^0 \rho_n^2 + (K_A + pV_A - TS_A)\rho_n + \frac{\kappa_A}{2}(\nabla \rho_n)^2 + \frac{\kappa_A'}{2}(\nabla^2 \rho_n)^2\right] d\vec{r} \quad (8)$$

where we use a three-dimensional notation, $\int (\nabla \rho_n)^2\,dV = -\int \rho_n \nabla^2 \rho_n\,dV$ and $\int (\nabla^2 \rho_n)^2\,dV = -\int \rho_n \nabla^4 \rho_n\,dV$ with $\nabla \rho_n = 0$ and $\nabla^3 \rho_n = 0$ at the boundaries of the integrals. Here two density-dependent terms $(K_A + pV_A - TS_A)\rho_n$ and $E_A^0 \rho_n^2$ appear in the free energy functional that scale differently with respect to the density $\rho_n$.

The bulk atomic densities on the two sides of the GB can have different isotropy, depending on the crystallographic planes that meet at the GB plane. This can result in asymmetric GB structure and density profile. In the current study, however, we neglect this possibility and assume the same bulk density on the two sides of the GB. Thus we can further simplify our description by defining a dimensionless density parameter $\rho(x) = \frac{\rho_n(x)}{\rho_n^B}$ with $\rho_n^B = \rho_n(-\infty)$ and writing

$$G_A = E_A^B \rho^2 + (K_A^B + pV_A^B - TS_A^B)\rho + \frac{\kappa_\rho}{2}(\nabla \rho)^2 + \frac{\kappa_\rho'}{2}(\nabla^2 \rho)^2 \quad (9)$$

in which $E_A^B = E_A^0 (\rho_n^B)^2$, $\kappa_\rho = \kappa_A (\rho_n^B)^2$, $\kappa_\rho' = \kappa_A' (\rho_n^B)^2$, $K_A^B = K_A \rho_n^B$, $pV_A^B = pV_A \rho_n^B$, $S_A^B = S_A \rho_n^B$ and, the Gibbs free energy of the homogeneous bulk phase (for $\rho = 1$) recovers as

$$G_A^B = E_A^B + K_A^B + pV_A^B - TS_A^B = H_A^B - TS_A^B. \tag{10}$$

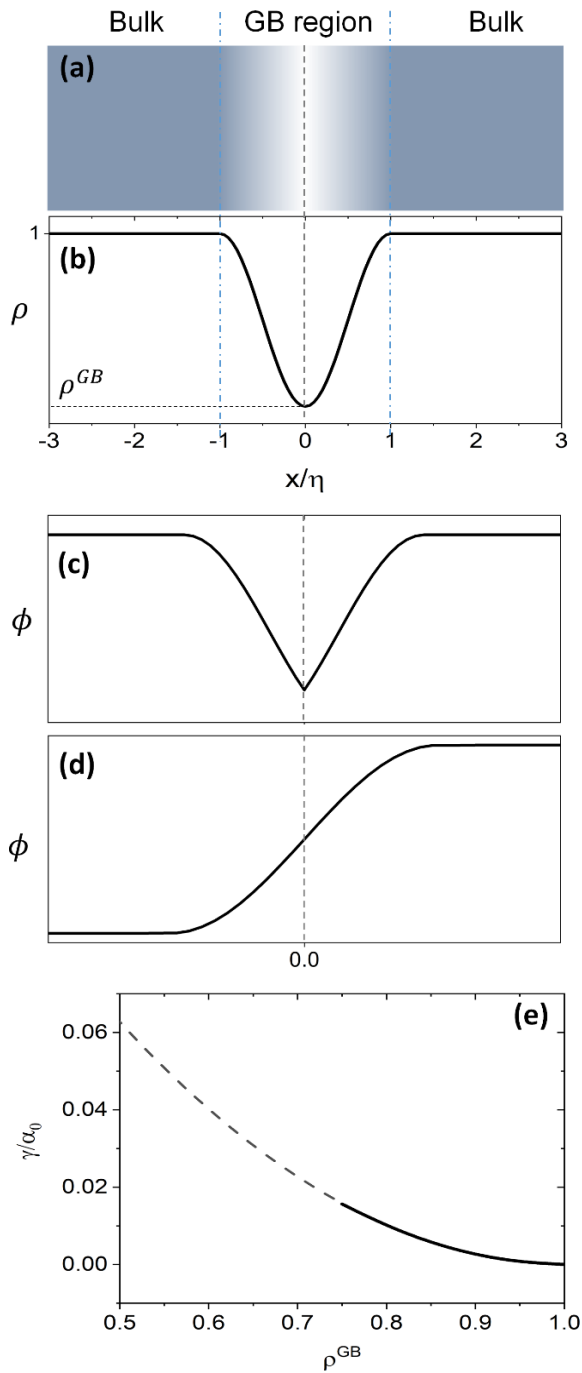

Figure 1: (a) and (b) Continuous density profile across a symmetric flat GB with a density $\rho(x = 0) = \rho^{GB}$ (Eq. (13)). The GB width is related to the model parameters $E_A$ and $\kappa_A$. The current density profile across GB is compared versus schematic drawing of order parameters from (c) the KWC phase-field model for GBs [47] usually applied for studying GB phase transitions [48-50] and (d) the classical phase-field profile (see for instance [65]). (e) The energy of the GB as a function of its density (Eq. (7)) is shown (see Sec. 4.2 for discussions).

To obtain the GB free energy, one can compare a heterogeneous system including a GB against a homogeneous bulk system (without the boundary). Thus, subtracting (10) from (9) we obtain

$$G_A^{GB} = E_A^B(\rho^2 - 1) + (K_A^B + pV_A^B - TS_A^B)(\rho - 1) + \frac{\kappa_\rho}{2}(\nabla \rho)^2 + \frac{\kappa_\rho'}{2}(\nabla^2 \rho)^2. \quad (11)$$

Equation (11) gives an approximation of the GB Gibbs free energy density, $G_A^{GB}$. For the symmetric flat GB centered at position $x = 0$, the equilibrium density profile across the GB region ($0 \leq x \leq \eta$) will be

$$\rho(x) = \rho^{GB} - \sum_{i=1}^{4} C_i + C_1 e^{\beta_1 x} + C_2 e^{-\beta_1 x} + C_3 e^{\beta_2 x} + C_4 e^{-\beta_2 x} \quad (12)$$

which satisfies $\frac{\delta G_A^{GB}}{\delta \rho} = 0$ with $\kappa_\rho' > 0$. The coefficients $C_i$s need to fulfill boundary conditions $\rho(x = 0) = \rho^{GB}$, $\rho(x = \eta) = 1$ and the continuity conditions $\frac{\partial \rho}{\partial x}(x = 0) = \frac{\partial \rho}{\partial x}(x = \eta) = 0$. Here $\beta_1 = \sqrt{\frac{\kappa_\rho - \sqrt{\kappa_\rho^2 - 8 E_A^B \kappa_\rho'}}{2\kappa_\rho'}}$, $\beta_2 = \sqrt{\frac{\kappa_\rho + \sqrt{\kappa_\rho^2 - 8 E_A^B \kappa_\rho'}}{2\kappa_\rho'}}$ and $\eta$ is the GB half width. For a simpler case, assuming $\kappa_\rho' = 0$, the equilibrium density profile across the GB region reads

$$\rho(x) = \left(\frac{1 + \rho^{GB}}{2}\right) - \left(\frac{1 - \rho^{GB}}{2}\right) \cos\left(\frac{\pi x}{\eta}\right) \quad (13)$$

with $\eta = \pi \sqrt{\frac{\kappa_A}{-2 E_A^B}}$. Equations (12) or (13) give a continuous atomic density profile across the GB. For more details on these equations see Appendix A. The continuity of atomic density profile across the GB region is confirmed by atomistic simulations [61]. Figure 1 compares the atomic density profile Eq. (13) versus previous phase-field models for GBs [47-50] (Fig. 1(c)) as well as the classical phase-field profile across an interface [65-67] (Fig. 1 (d)). Using Eq. (13) with $\kappa_\rho' = 0$, the GB energy $\gamma = 2 \int_0^\eta G_A^{GB} dr$ can be analytically obtained as

$$\gamma = \alpha_0 (1 - \rho^{GB})^2 \quad (14)$$

with $\alpha_0 = \frac{\pi}{4} \sqrt{-2 E_A^B \kappa_A}$. This equation approximates the direct relation between $\gamma$ and $\rho^{GB}$ for an equilibrium density profile across the GB. Figure 1 (e) shows the normalized value $\frac{\gamma}{\alpha_0}$ as a function of $\rho^{GB}$. In the following, $\rho^{GB}$, the average relative atomic density within the GB plane, will be referred to as 'GB density'. The significance of the GB density and its relationship with the GB character is discussed in Sec. 4.2.

## 2.2 A GB in a regular binary alloy

In order to extend the current density-based model for a GB in a binary alloy, we need to discuss the significance of mixing energy as a function of atomic density. For a bulk regular solution made of A (solvent) and B (solute) atoms, the change in the Gibbs free energy due to the mixing can be approximated as [68]

$$\Delta G_{mix}^B(X_B) = \Delta H_{mix}^B - T\Delta S_{mix}^B = \Omega\, X_A X_B + RT[X_A \ln X_A + X_B \ln X_B] \tag{15}$$

with

$$\Omega = N_a\, Z\, \Delta\epsilon \tag{16}$$

in which $R$ is the gas constant, $\Omega$ is the mixing enthalpy coefficient, $N_a$ is Avogadro number and $Z$ is the coordination number. $\Delta\epsilon = \left(\epsilon_{AB} - \frac{\epsilon_{AA}+\epsilon_{BB}}{2}\right)$ is bonding energy in which $\epsilon_{ij}$ is the bonding energy between atoms $i$ and $j$ and $X_A$ ($X_B$) is the mole fraction of atoms $A$ ($B$). In principle, the mixing energy terms in Eq. (15) can be assumed to follow the same scaling found in Eq. (9). The effect of changing the coordination number $Z$ on the GB enthalpy of mixing and GB segregation has been previously discussed (see for instance [40, 46, 69] and references therein). While the coordination number is proportional to the local density, the stretch in the bonding energy $\Delta\epsilon$ is also expected to change with the density in the limit of linear elasticity. Considering these facts and neglecting the dependency of configurational entropy for the sake of simplicity, we can write

$$\Delta G_{mix}(X_B, \rho) \approx \rho^2 \Delta H_{mix}^B - T\Delta S_{mix}^B. \tag{17}$$

Following Cahn and Hilliard [44], an energy term due to the concentration gradients –in the presence of chemical heterogeneity– enters the free energy description as well. Thus, using Eqs. (9) and (17), the density-based Gibbs free energy of a regular solid solution (SS) can be written as

$$\begin{aligned}
G_{SS}(X_B, \rho) &= X_A G_A(\rho) + X_B G_B(\rho) + \Delta G_{mix}(X_B, \rho) \\
&= X_A\left(E_A^B \rho^2 + (K_A^B + pV_A^B - TS_A^B)\rho + \frac{\kappa_A}{2}(\nabla\rho)^2 + \frac{\kappa'_\rho}{2}(\nabla^2\rho)^2\right) \\
&\quad + X_B\left(E_B^B \rho^2 + (K_B^B + pV_B^B - TS_B^B)\rho + \frac{\kappa_B}{2}(\nabla\rho)^2 + \frac{\kappa'_\rho}{2}(\nabla^2\rho)^2\right) \\
&\quad + \rho^2 \Omega\, X_A\, X_B - T\Delta S_{mix}^B + \frac{\kappa_X}{2}(\nabla X_B)^2
\end{aligned} \tag{18}$$

where $\kappa_X$ is the concentration gradient coefficient. In Eq. (18), corresponding molar fractions $X_i^B$ and $X_i^{GB}$ will be considered for the bulk and GB regions, respectively. For the homogeneous bulk phase, one recovers

$$G_{SS}^B(X_B^B, \rho = 1) = X_A^B(H_A^B - TS_A^B) + X_B^B(H_B^B - TS_B^B) + \Omega\, X_A^B X_B^B - T\Delta S_{mix}^B. \qquad (19)$$

Using Eq. (18) we are able now to study the GB phase equilibria and segregation isotherms as presented in the next section.

## 3. Applications of the Current Density-based Model

### 3.1 Equilibrium GB phase diagram

Obviously, a GB may not exist without its corresponding bulk phase(s). It is, however, useful to study the equilibrium phase diagram of a hypothetical GB as a function of its atomic density. As Eq. (18) suggests, for such hypothetical GB with $\rho < 1$ the phase equilibria should differ from that of the corresponding bulk material with $\rho = 1$. The equilibrium states are evaluated by minimizing the Gibbs free energy functional with respect to the concentration and atomic density fields, i.e. $\frac{\delta \mathcal{G}_{alloy}}{\delta \rho} \to 0$ and $\frac{\delta \mathcal{G}_{alloy}}{\delta X_B} \to 0$ with $\frac{\delta \mathcal{G}_{alloy}}{\delta q} = \frac{\partial \mathcal{G}_{alloy}}{\partial q} - \nabla \frac{\partial \mathcal{G}_{alloy}}{\partial \nabla q} + \nabla^2 \frac{\partial \mathcal{G}_{alloy}}{\partial \nabla^2 q}$ and $\mathcal{G}_{alloy} = \int G_{SS}\, d\vec{r}$. A numerical solution of these equations can provide information about the spatiotemporal evolution of phases in the GB region. This has been quantitatively demonstrated in a parallel study on the Mn segregation in Fe-Mn system [61]. The aim of the current study, however, is to provide analytical descriptions for equilibrium GB thermodynamics. Hence, in the following, we rather investigate a single material point at the GB center where spatial gradients of the atomic density field vanish, i.e. $\nabla \rho(x = 0) = 0$. We further assume that for this point the concentration gradient and higher-order density gradients can be neglected. Thus, with $\rho(x = 0) = \rho^{GB}$ and $X_B(x = 0) = X_B^{GB}$, the density-based Gibbs free energy description at this material point simplifies to

$$\begin{aligned}G_{SS}(X_B^{GB}, \rho^{GB}) = &\; X_A^{GB}(E_A^B(\rho^{GB})^2 + (K_A^B + pV_A^B - TS_A^B)\rho^{GB}) \\ &+ X_B^{GB}(E_B^B(\rho^{GB})^2 + (K_B^B + pV_B^B - TS_B^B)\rho^{GB}) \\ &+ (\rho^{GB})^2 \Omega\, X_A^{GB} X_B^{GB} - T\Delta S_{mix}^B.\end{aligned} \qquad (20)$$

Equation (20) allows to approximate the GB thermodynamic properties, analytically. To demonstrate application of the current density-based model, we consider a hypothetical GB with $\rho^{GB} = 0.75$ in the Pt-Au system. In this alloy system, the only difference to the theoretical regular solution is in the enthalpy of mixing that extends to the second term in Redlich-Kister polynomial:

$$\Omega_{\text{PtAu}} = L_0 + L_1(X_{Pt}^B - X_{Au}^B). \tag{21}$$

Inserting Eq. (21) in Eq. (20) we obtain

$$\begin{aligned} G_{SS}(X_{Au}^{GB}, \rho^{GB}) = {} & X_{Pt}^{GB}\left(E_{Pt}^B(\rho^{GB})^2 + (K_{Pt}^B + pV_{Pt}^B - TS_{Pt}^B)\rho^{GB}\right) \\ & + X_{Au}^{GB}\left(E_{Au}^B(\rho^{GB})^2 + (K_{Au}^B + pV_{Au}^B - TS_{Au}^B)\rho^{GB}\right) \\ & + (\rho^{GB})^2 \Omega_{\text{PtAu}} X_{Pt}^{GB} X_{Au}^{GB} - T\Delta S_{mix}^B \,. \end{aligned} \tag{22}$$

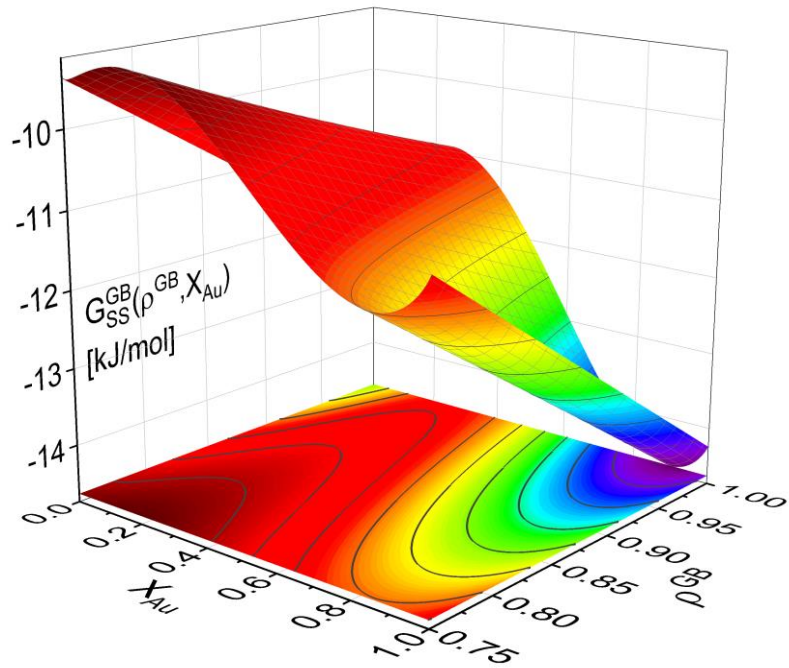

Figure 2: Gibbs free energy density of the Pt-Au system (Eq. (22)) is shown as a generalized function of composition and atomic density at T = 300 K. For $\rho^{GB} = 1 = \rho^B$, the Gibbs free energy of the reference bulk is recovered. For $\rho^{GB} < 1$ the free energy increases.

Grolier *et al*. [70] have assessed the bulk thermodynamic data for binary the Pt-Au and reported $L_0 = 11625 + 8.3104\,T$ and $L_1 = -12616 + 5.8186\,T$ [J/mol] for the FCC Pt-Au solid solution. The rest of the thermodynamic data for pure Pt and Au are extracted from SGTE compilation by Dinsdale [71]. Inserting these values in Eqs. (21) and (22), one obtains the Gibbs free energy as a function of GB density and composition. Figure 2 shows a 3D map of the Gibbs free energy for the Pt-Au system at 300 K as a function of composition $X_{Au}$ and GB density $\rho^{GB}$. We found that as the GB density decreases, the Gibbs free energy increases and deviates larger from the corresponding bulk values. Figure 3 compares bulk and GB Gibbs free energy for $\rho^{GB} = 0.75$ at different temperatures.

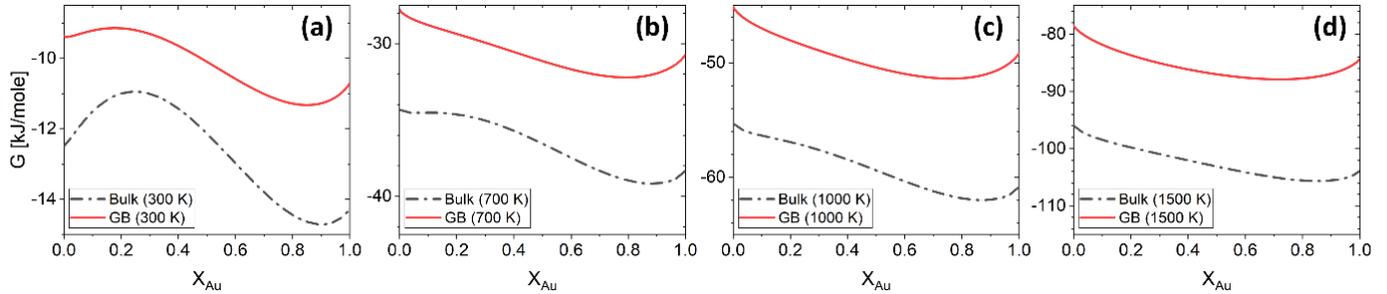

Figure 3: The Gibbs free energy density of the bulk and our hypothetical GB with $\rho^{GB} = 0.75$ in the Pt-Au system obtained from Eq. (22) are compared at different temperatures. As expected, the GB Gibbs free energy density is higher than the bulk.

Using Eq. (22), the equilibrium phase diagrams of the bulk and any specific standalone GB (with a given GB density) in the Pt-Au system can be generated. Figure 4 compares the phase diagrams of the bulk and our hypothetical GB (with $\rho^{GB} = 0.75$) in the Pt-Au solid solution. The results show that, inside a (hypothetically standalone) GB, the interfacial phase separation becomes possible but for smaller ranges of temperature and composition when compared to the bulk phase diagram.

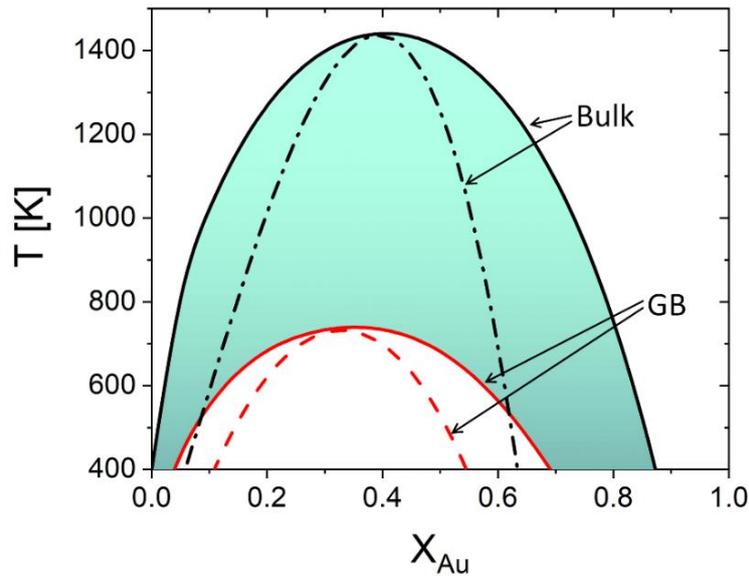

Figure 4: Equilibrium phase diagram for the bulk $\rho = \rho^B = 1$ and our hypothetical GB with $\rho = \rho^{GB} = 0.75$ are obtained for the Pt-Au system. The dash lines represent the chemical spinodals. The colored area marks the possible domain where GB phase diagrams for $0.75 < \rho^{GB} < 1$ can appear. Note that the equilibrium bulk and GB phase diagrams are plotted independently. The coexistence of the bulk and GB phases will be discussed in Secs. 3.2 and 4.1 and Fig. 6.

As it is evident from the Gibbs free energy plots (Figs. 2 and 3), the GB thermodynamics strongly depend on the GB density $\rho^{GB}$. Although the phase diagram is a nonlinear function of the GB density, the possible values for GB density $\rho^{GB}$ are finite and vary close to 1. For instance, the colored area in Fig. 4 indicates where the GB phase diagram can appear for $0.75 < \rho^{GB} < 1$. Thus, once the range of GB density is determined, the equilibrium GB phase diagrams can be approximated. The GB density represents the nature (type) of the GB and correlates with the GB energy and misorientation angle. These aspects of the current density-based model are discussed in Sec. 4.2.

**3.2 GB segregation behavior**

In Section 3.1 a hypothetical standalone GB was discussed. For a complete description of GB phase equilibria, however, the coexistence between the GB and bulk phases must be understood. In a binary system, this requires equality of the (relative) chemical potentials $\mu_B^B - \mu_A^B = \mu_B^{GB} - \mu_A^{GB}$ all across the system (parallel tangent construction) [24, 72]. In reality, these are not sufficient conditions as a minimum energy state with respect to the atomic density field must be satisfied as well. However, if we consider the same material point at the GB center as above, we can simplify the problem and solve for the equality of the (relative) chemical potentials using Eq. (20) which gives

$$\frac{X_B^{GB}}{1 - X_B^{GB}} = \frac{X_B^B}{1 - X_B^B} \cdot \exp\left(-\frac{[\Delta E^B + \Omega]\left(\rho^{GB^2} - 1\right) + (\Delta K^B + p\Delta V^B - T\Delta S^B)(\rho^{GB} - 1) + 2\Omega\left[X_B^B - \rho^{GB^2} X_B^{GB}\right]}{RT}\right). \quad (23)$$

Here $X_B^{GB}$ is the composition of the GB with $\rho = \rho^{GB}$, $X_B^B$ is the composition of the bulk far from the GB, $\Delta E^B = E_B^B - E_A^B$, $\Delta K^B = K_B^B - K_A^B$, $p\Delta V^B = pV_B^B - pV_A^B$, and $\Delta S = S_B^B - S_A^B$. Equation (23) resembles the Fowler-Guggenheim segregation isotherm [24] but also takes the specific effect of a given GB, represented by its density $\rho^{GB}$, into account. For the Pt-Au system, we obtain

$$\frac{X_{Au}^{GB}}{1 - X_{Au}^{GB}} = \frac{X_{Au}^B}{1 - X_{Au}^B} \cdot \exp\left(-\frac{[\Delta E^B + L_0]\left(\rho^{GB^2} - 1\right) + (\Delta K^B + p\Delta V^B - T\Delta S^B)(\rho^{GB} - 1) + 2L_0\left[X_{Au}^B - \rho^{GB^2} X_{Au}^{GB}\right] + Q}{RT}\right) \quad (24)$$

with
$$Q = L_1\left[2\rho^{GB^2}X_{Au}^{GB}(1-X_{Au}^{GB}) - (1-2X_{Au}^{GB})^2 - 2X_{Au}^B(1-X_{Au}^B) + (1-2X_{Au}^B)^2\right],$$

$\Delta E^B = E_{Au}^B - E_{Pt}^B$, $\Delta K^B = K_{Au}^B - K_{Pt}^B$, $p\Delta V^B = pV_{Au}^B - pV_{Pt}^B$, and $\Delta S = S_{Au}^B - S_{Pt}^B$. Equations (23) and (24) use GB density and bulk thermodynamic data as the inputs and give the corresponding GB segregation isotherms.

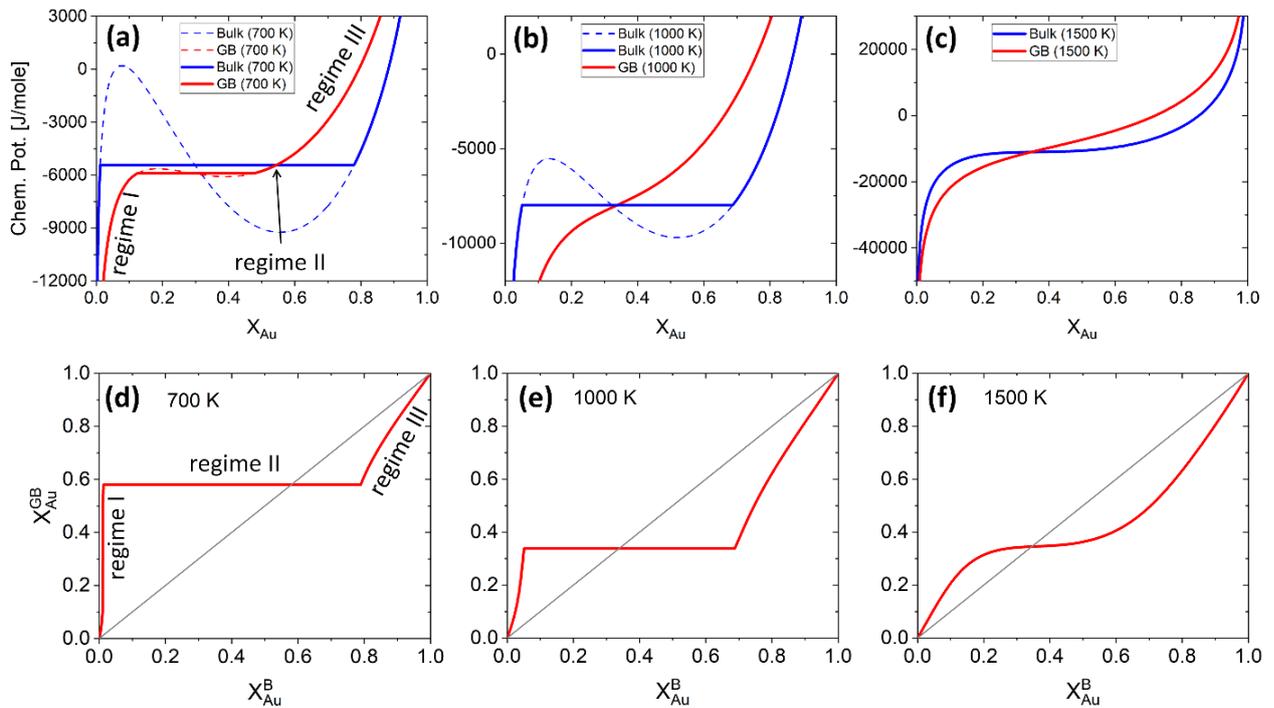

Figure 5: (a)-(c) The chemical potentials of the Au atoms in bulk and a GB (with $\rho^{GB} = 0.75$) Pt-Au are plotted for three different temperatures. (d)-(f) Using the parallel tangent condition the segregation isotherms are obtained for the corresponding temperatures. Different regimes of segregation appear depending on the GB density and temperature. See discussions in Sec. 4.2.

Figure 5 presents the (relative) chemical potentials and the segregation isotherms for the bulk and the exemplar GB with $\rho^{GB} = 0.75$ in the Pt-Au system at three different temperatures. The chemical potential of the GB which is a function of GB density differs from the corresponding bulk chemical potential. The Maxwell construction (defining the two-phase regions) and the spinodal compositions for both GB and bulk materials are built. The graphs show that the GB spinodal decomposition (interfacial spinodal) can occur for lower chemical potential values below the bulk spinodal (e.g. in Fig. 5 (a)). As a result, the two-phase GB can be in equilibrium with a single-phase bulk. As a matter of fact, it has been reported that a segregation—assisted interfacial phase separation can occur in Pt-Au system which results in the formation of two-phase GBs coexisting with a single-phase bulk [73, 74]. The results in Fig. 5 suggest that the two-phase bulk will be in equilibrium with a single-phase GB. In fact, at a given temperature, for any bulk composition within the two-phase region of the phase diagram, the two-phase bulk is expected to be in equilibrium with a single phase/composition GB. These are because of a vertical shift in the GB chemical potential (with respect to the bulk chemical potential) that can be seen in Fig. 5 (a). Combining these results with the equilibrium bulk and GB phase diagrams (Sec. 3.1), we are able now to study the complete GB thermodynamic phase diagram. This is further discussed in Sec. 4.

## 4. Discussion

### 4.1 Coexistence of GB and bulk phases

In this model, the GB region is described by a continuous atomic density and its spatial gradients. For the sake of simplicity, here we have limited our derivations to the first-order approximation of the potential energy with respect to $\rho$ and its special variations. The current density-based model provides a simple method for approximating the Gibbs free energy of a given GB based on available bulk thermodynamic data. The continuity of the atomic density field across GB allows simplifying the Gibbs free energy description at the GB plane where the gradient term $\nabla \rho$ vanishes. The results of the current study are obtained for this single material point within the GB region. The model can be applied in full-field simulations of GBs as well, to evaluate the temporal evolution of the atomic density and concentration fields. In two parallel study, we have demonstrated the application of the density-based model for studying GB segregation in a FeMnNiCrCo high entropy alloy [60] and segregation engineering of Fe-Mn alloys [61].

In the current study, we focus on the equilibrium GB properties. Using the density-based formulation we extend the concept of the phase diagram for GBs for binary solid solutions, Fig. 6 (b). We started with the equilibrium phase diagram for the bulk and a hypothetical standalone GB, with $\rho^{GB} = 0.75$, in the Pt-Au solid solution (Figure 4). The results show that like the bulk materials, GBs in the Pt-Au system can also undergo an interfacial spinodal phase separation. In this case, we show that GB two-phase region is confined to smaller temperature and composition ranges. As the GB density deviates larger from the bulk density, the GB miscibility gap in the phase diagram becomes smaller.

The coexistence of the bulk and GB phases can be studied using the equal chemical potential condition [24, 72]. Here the bulk phase, which is the dominant part of the microstructure, dictates the ultimate thermodynamic states and behavior of the GBs. According to Eq. (23) (or Eq. (24) for the Pt-Au system), the GB segregation level depends on the temperature, bulk composition, as well as GB type represented by its density $\rho^{GB}$. For a regular solution discussed here, the driving forces for the segregation can be divided into two parts: The first contribution is the ideal segregation energy $\Delta E^B \left( \rho^{GB^2} - 1 \right) + (\Delta K^B + p\Delta V^B - T\Delta S^B)(\rho^{GB} - 1)$ that is analogous to the segregation driving force in Langmuir-McLean isotherm [72]. If we neglect the energetic contributions due to the mixing ($\Omega = 0$), we get a density-based version of Langmuir-McLean relation as:

$$\frac{X_B^{GB}}{1-X_B^{GB}} = \frac{X_B^B}{1-X_B^B} \cdot \exp\left(-\frac{\Delta E^B\left(\rho^{GB\,2}-1\right) + (\Delta K^B + p\Delta V^B - T\Delta S^B)(\rho^{GB}-1)}{RT}\right) \qquad (25)$$

that describes solute segregation for a given GB in an ideal solution. The second contribution in Eq. (23) is $\Omega\left(\rho^{GB\,2}-1\right) + 2\Omega\left[X_B^B - \rho^{GB\,2}X_B^{GB}\right]$ due to the mixing which contains a term with the GB concentration $X_B^{GB}$. The two contributions discussed above can have cooperative or competitive effects on the segregation level, depending on the magnitude and sign of $\Omega$ and $\Delta E^B$.

Figure 5 presents the chemical potentials and segregation isotherms for the Pt-Au system. In this system, three regimes of bulk-GB coexistence could be identified as a function of the bulk composition. First, segregation to the GB occurs as the Au content increases in Pt-rich alloys (regime I). For a two-phase bulk, the GB composition becomes fixed (regime II) that applies to the entire range of bulk composition in the two-phase bulk region. Any further increase of the Au content results in solute depletion in the GB (regime III) with respect to the bulk. In Figs. 5 (a) and (d) different regimes of segregation are marked.

For a Pt-rich Pt-Au alloy (regime I), the GB, which is enriched in Au due to the segregation, can undergo an interfacial spinodal decomposition before a bulk spinodal becomes possible. At 700 K, for instance, a jump in the GB segregation, associated with the interfacial phase separation, is revealed in the Pt-Au system. Figure 6 (a) shows this jump in the GB composition as a function of bulk composition (this is a zoomed-in plot of the left corner in Fig. 5 (d)). At this jump, the GB decomposes into low-concentration and high-concertation domains that occurs in the presence of a single-phase bulk. This means that for a specific bulk composition (in the single-phase region of the bulk phase diagram) a spinodally-decomposed two-phase GB will form. Consistent with the current predations, a segregation—assisted interfacial spinodal decomposition has been indeed reported in nanocrystalline Pt-Au alloy [73]. Also, atomistic simulations of the Pt-Au system confirmed such interfacial spinodal phase separation for bulk compositions well below the bulk spinodal range [74]. GB spinodal phase separation is also evidenced in other material systems with technological significance [75, 76].

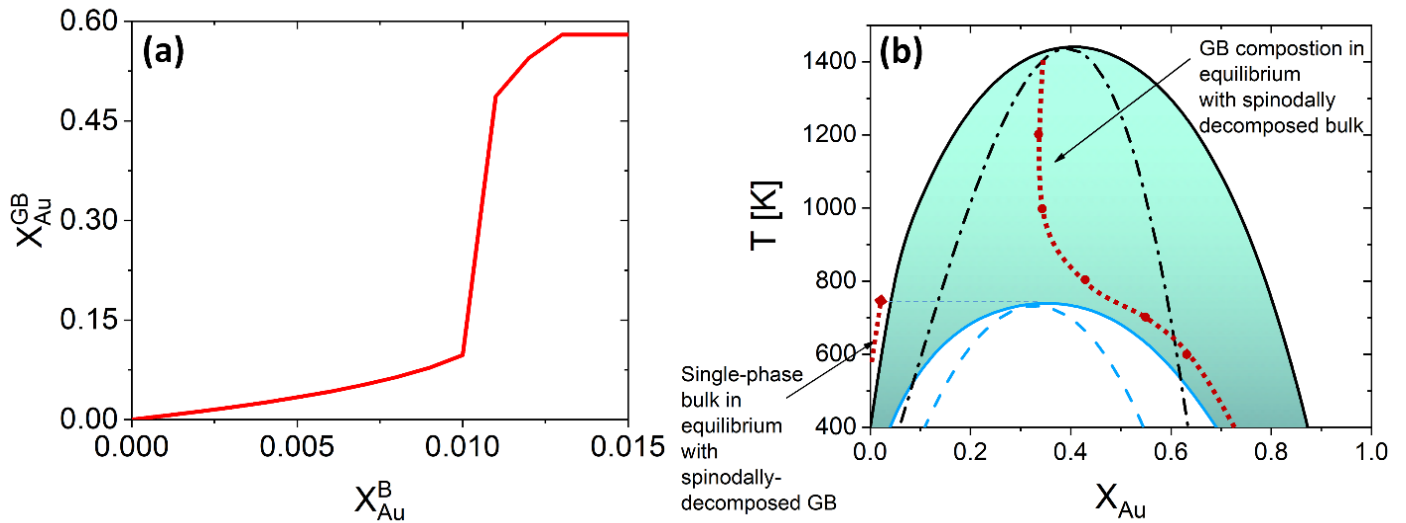

Figure 6: (a) For Pt-Au system at 700 K, a jump (GB spinodal) occurs before the bulk spinodal decomposition takes place. This is a zoomed in plot in Fig. 5 (d) (left corner), see also Fig. 5 (a).
(b) Based on the current density-based model the coexistence of the bulk and GB phases is depicted in the phase diagram for the Pt-Au system. A two-phase bulk will be in equilibrium with a single-phase GB while a two-phase (spinodally-decomposed) GB is shown to be in equilibrium with a single-phase bulk.

For the two-phase Pt-Au bulk materials (in regime II), the model predicts that a single-phase GB comes to coexist. This is because of the vertical shift in the GB chemical potential with respect to the corresponding bulk chemical potential curve, as depicted in Fig. 5 (a). In practice, a range of atomic densities and compositions exists in the GB region (between the GB plane and bulk interior) that can enable interfacial phase separation in the GB region. Combining the equilibrium GB phase diagram (Fig. (4)) and the GB segregation isotherms (Fig. (5)), we can obtain the complete GB phase diagram as presented in Fig. 6 (b).

The discontinuous jump in the GB segregation isotherm (Fig. 6 (a)) is a result of the difference in the free energy functional and chemical potentials of the bulk and GB phases, captured by the density-based model. Depending on the sign of $\Delta E^B$ and the shape of $\Omega(X_B)$ function (in composition space), the GB chemical potential and its corresponding spinodal decomposition can lie 'Above' or 'Below' the bulk spinodal (Figure 7). In either case, one can activate a segregation—assisted interfacial phase separation, that can act as a precursor for subsequent nucleation, before reaching the bulk two-phase region. It has been shown that a segregation—assisted *transient* spinodal phase separation comes to exist at GBs for a larger range of bulk compositions [61] which enables manipulation of GB segregation by desired heat treatments.

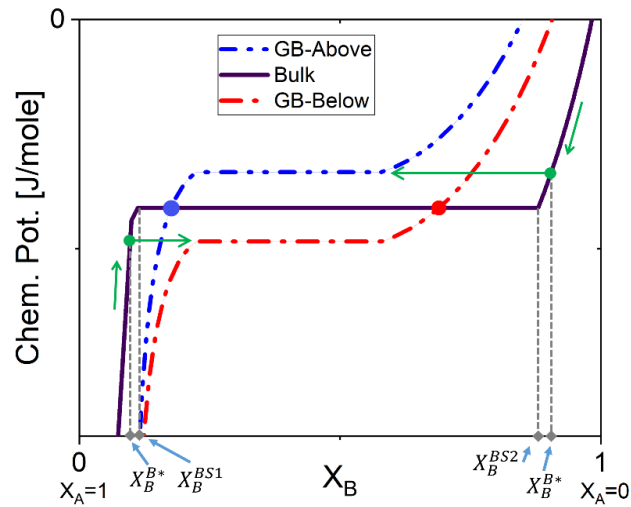

Figure 7: The GB chemical potential can be 'Below' or 'Above' the bulk's curve. $X_B^{BS1}$ and $X_B^{BS2}$ show the bulk equilibrium compositions in the two-phase region. The two-phase GB will be in equilibrium with a single-phase bulk material as marked by the arrows. The dots on the bulk chemical potential curve indicates the single-phase GB is equilibrium with the two-phase bulk.

In a polycrystalline material having a large population of GBs of various types (with different crystallographic properties), the GB densities are expected to differ from one GB to another. A quantitative application of the current model to technical materials, therefore, requires determination of the GB structure and densities for different types of GBs. For this purpose, atomistic simulations can provide the necessary means to evaluate GB density profiles. Though atomistic simulation of GBs is beyond the scope of this study, in the following we briefly discuss the relationships between the $\rho^{GB}$, GB excess free volume and GB misorientation angle that can provide more insights to the density-based concept.

### 4.2 The relation between the GB density and GB nature

In order to make use of the current model, it is helpful to explore the significance of the average GB density and its relationship with the GB nature. Using atomistic simulation of any GB, 3D density map can be extracted from the equilibrium atomic configuration over a physical coarse-graining radius that gives $\eta$ and $\rho^{GB}$ [61]. GB energy $\gamma$ can be related to the $\eta$ and $\rho^{GB}$ values. Eq. (14) approximates the relationship between GB energy and its density $\rho^{GB}$ when $\kappa'_\rho = 0$ is assumed. It is observed that GBs with higher energies attract more solute content during the segregation [77]. Using Eqs. (13) and (18), the GB Gibbs free energy of a regular solution can be written as a function of its initial GB energy. Accordingly, we can rewrite Eq. (23) as

$$\frac{X_B^{GB}}{1-X_B^{GB}} = \frac{X_B^B}{1-X_B^B} \cdot \exp\left(-\frac{[\Delta E^B + \Omega]\left(\left[1-\sqrt{\frac{\gamma}{\alpha_0}}\right]^2 - 1\right) + (\Delta K^B + p\Delta V^B - T\Delta S^B)\left(\left[1-\sqrt{\frac{\gamma}{\alpha_0}}\right] - 1\right) + 2\Omega\left[X_B^B - \left[1-\sqrt{\frac{\gamma}{\alpha_0}}\right]^2 X_B^{GB}\right]}{RT}\right) \quad (26)$$

that gives a useful approximation for describing GB's tendency for segregation as a function of its initial energy.

One way to elaborate about the GB density $\rho^{GB}$ is to find its relationship with the GB excess free volume defined as [78]

$$\Delta V = \frac{WN^B}{N_0}\left(1 - \frac{N^{GB}}{N^B}\right). \quad (27)$$

Here $W$ is the atom diameter, $N^{GB}$ is the number of atoms in the GB, $N^B$ is the number of atoms in the bulk within the same volume and $N_0$ is the number of atoms per unit area of the GB. In principle, the 'excess' volume represents the 'shortage' of atoms in the GB plane. On the other hand, comparing a GB against the bulk of the same volume and using Eq. (13), the relative number of atoms in a GB will be $n^{GB} = \frac{(1+\rho^{GB})}{2}$ (where $n^B = 1$) and the relative difference (with respect to the bulk) will be

$$\Delta n = 1 - \rho^{GB} \quad (28)$$

where $n = 2\int_0^\eta \rho\, dr$ is the (relative) total number of atoms in the GB. Comparing Eqs. (27) and (28) one can find the one-to-one relationship between $\Delta n$ and $\Delta V$, where we can identify $\Delta V \propto (1 - \rho^{GB})$. Extensive works have been devoted to calculate and measure GB excess free volume (see for instance [79-81]). Aaron and Bolling [78] have shown that the energy of different types of GBs can be associated with their excess free volume which can be extended as well for the GB density parameter $\rho^{GB}$.

The lower limit of the GB density value (highest excess free volume) shall be obtained for general high angle GBs (HAGBs). Arron and Bolling remarked that the excess free volume in a general HAGB is comparable to that of a liquid phase [78]. On the other hand, the maximum GB density value is the corresponding bulk density $\rho^B = 1$. Special GBs such as coherent twin boundaries (TBs) with low coincidence values show small GB energy and excess free volumes [78]. This is well consistent with the current model in which as $\Delta V$ goes to zero, $\rho^{GB}$ approaches $\rho^B = 1$ and $\gamma$ approaches zero (Eq. (14)).

For a low-angle GB (LAGB), the situation becomes a bit more complex because of the dislocations. Since dislocations are localized defects with a lower atomic density in their core, one expects the local GB density $\rho^{GB}$ within the LAGB plane to fluctuate accordingly. In addition, elastic interactions between dislocations can play a role in determining $\rho^{GB}$ for LAGBs. For a simple case neglecting these elastic interactions, we can approximate the average GB density for a LAGB by a simple volume averaging of regularly spaced dislocations, as detailed in Appendix B. For a tilt GB, a simple geometrical construction of a LAGB gives the average GB density $\rho^{GB}_{tilt} \approx 1 - \frac{\sin\theta}{4}$ with the GB misorientation $\theta$, and the edge dislocation core density $\rho^D \approx 0.75$. For a larger misorientation angle, the GB density will be smaller, deviating larger from bulk properties. We also obtain a misorientation-dependent $\Delta n \approx \frac{\sin\theta}{4}$ and thus a corresponding GB excess free volume $\Delta V \propto \sin\theta$ that is consistent with the previous studies [78, 82]. GB segregation as a function of GB misorientation has been studied in the Pt-Au system. Seki *et al.* [83, 84] and Seidman [85] studied Au segregation to twist boundaries and have shown that although the GB segregation is not homogeneous, its average level increases as the misorientation angle increases (with the exception of special GBs). These observations confirm the results from the current density-based model where a higher misorientation angle corresponds to a lower GB density. An in-depth study of LAGBs requires consideration of elastic energy contributions which is left for a future study.

## 5. Summary and outlook

In this work, we have derived a density-based model for assessing GB thermodynamics. The current model uses available bulk thermodynamic data, as input, and gives a rather general formulation for GB's Gibbs free energy. GB segregation isotherm and phase diagram for a regular binary solid solution have been obtained. We further discussed the relationships between GB density, GB excess free volume and, GB misorientation angle. The results are qualitatively demonstrated on the Pt-Au system. We show that in a Pt-rich Pt-Au alloy, Au atoms tend to segregate to the GB. The average level of segregation increases with the GB misorientation angle (with the exception of special GBs). We also show that an interfacial spinodal decomposition can occur in these alloys for single-phase bulk alloys. A comparison with previous studies on the Pt-Au system confirms these predictions.

The current density-based model offers a simple, pragmatic approach to develop GB thermodynamic databases. Despite its simplicity, applications of this model can unroll new microstructure design concepts by segregation engineering of GBs [60, 61]. A systematic atomistic study of the atomic density profiles for different types of GBs can enrich the development of this model. In order to assess engineering alloys with interstitial solute atoms, the density-based free energy description needs to be extended for multi-component materials and including elastic energy contributions.

## Appendix A: Equilibrium GB Density Profile

The equilibrium GB density profile across a symmetric flat GB in a pure substance can be obtained by minimizing the GB free energy functional with respect to the relative atomic density. For a 1D setup,

$$\frac{\delta \mathcal{G}_A}{\delta \rho} = \frac{\partial \mathcal{G}_A}{\partial \rho} - \frac{d}{dx}\left(\frac{\partial \mathcal{G}_A}{\partial \left(\frac{\partial \rho}{\partial x}\right)}\right) + \frac{d^2}{d^2 x}\left(\frac{\partial \mathcal{G}_A}{\partial \left(\frac{\partial^2 \rho}{\partial x^2}\right)}\right) = 2E_A^B \rho + C_A^B - \kappa_\rho \frac{\partial^2 \rho}{\partial x^2} + \kappa_\rho' \frac{\partial^4 \rho}{\partial x^4} = 0 \qquad \text{A.1}$$

where $C_A^B = K_A^B + pV_A^B - TS_A^B$ (see Eq. (9)). The general solution for this higher-order linear ordinary differential equation reads

$$\rho(x) = -\frac{C_A^B}{2E_A^B} + C_1 e^{\beta_1 x} + C_2 e^{-\beta_1 x} + C_3 e^{\beta_2 x} + C_4 e^{-\beta_2 x} \qquad \text{A.2}$$

with $\beta_1 = \sqrt{\frac{\kappa_\rho - \sqrt{\kappa_\rho^2 - 8E_A^B \kappa_\rho'}}{2\kappa_\rho'}}$ and $\beta_2 = \sqrt{\frac{\kappa_\rho + \sqrt{\kappa_\rho^2 - 8E_A^B \kappa_\rho'}}{2\kappa_\rho'}}$. We also obtain

$$\frac{\partial \rho(x)}{\partial x} = C_1 \beta_1 e^{\beta_1 x} - C_2 \beta_1 e^{-\beta_1 x} + C_3 \beta_2 e^{\beta_2 x} - C_4 \beta_2 e^{-\beta_2 x}. \qquad \text{A.3}$$

Applying the boundary conditions $\rho(x = 0) = \rho^{GB}$, $\rho(x = \eta) = 1$ and the continuity conditions $\frac{\partial \rho}{\partial x}(x = 0) = 0$ and $\frac{\partial \rho}{\partial x}(x = \eta) = 0$, we obtain

$$\rho^{GB} + \frac{C_A^B}{2E_A^B} = C_1 + C_2 + C_3 + C_4 \qquad \text{A.4}$$

$$1 + \frac{C_A^B}{2E_A^B} = C_1 e^{\beta_1 \eta} + C_2 e^{-\beta_1 \eta} + C_3 e^{\beta_2 \eta} + C_4 e^{-\beta_2 \eta} \qquad \text{A.5}$$

$$\frac{C_1 - C_2}{C_3 - C_4} = -\frac{\beta_2}{\beta_1} \qquad \text{A.6}$$

$$\frac{C_1 e^{\beta_1 \eta} - C_2 e^{-\beta_1 \eta}}{C_3 e^{\beta_2 \eta} - C_4 e^{-\beta_2 \eta}} = -\frac{\beta_2}{\beta_1} \qquad \text{A.7}$$

The solution to Eqs. A.4 to A.7 can be computed numerically. The continuity of the average atomic density profile across the GB region is confirmed using atomistic simulations [61]. For a simpler case assuming $\kappa'_\rho = 0$, we find another general solution for Eq. A.1:

$$\rho(x) = -\frac{C_A^B}{2E_A^B} + D_1 \cos\left(\sqrt{\frac{-2E_A^B}{\kappa_\rho}} x\right) + D_2 \sin\left(\sqrt{\frac{-2E_A^B}{\kappa_\rho}} x\right) \qquad \text{A.8}$$

where $D_1$ and $D_2$ are constants and

$$\frac{\partial \rho}{\partial x} = \sqrt{\frac{-2E_A^B}{\kappa_\rho}} \left[ -D_1 \sin\left(\sqrt{\frac{-2E_A^B}{\kappa_\rho}} x\right) + D_2 \cos\left(\sqrt{\frac{-2E_A^B}{\kappa_\rho}} x\right) \right] \qquad \text{A.9}$$

Applying the boundary conditions to this solution we obtain

$$D_1 = -\frac{1 - \rho^{GB}}{2} \qquad \text{A.10}$$

$$D_2 = 0 \qquad \text{A.11}$$

$$\eta = \pi \sqrt{\frac{\kappa_A}{-E_A^B}} \qquad \text{A.12}$$

$$\frac{C_A^B}{E_A^B} = -\frac{1 + \rho^{GB}}{2} \qquad \text{A.13}$$

Although solution Eq. A.8 is limiting, it provides a possibility for further analytical treatment of the density profile and therefore is used for approximating GB energy, Eq. (14). For more details see Sec. 2.1.

# Appendix B: A Simple Geometrical Calculation of GB Density

The objective of this appendix is to provide a simple analysis of the GB density and its dependence on the misorientation angle. We consider a LAGB that can be mapped with a set of regularly spaced dislocations separated by a distance [68]

$$D = \frac{b}{\sin\theta} \qquad \text{B.1}$$

in which $b$ is the Burgers vector of the dislocations and $\theta$ is the GB misorientation angle. The 'average' relative GB density can be obtained by averaging the density of a dislocation in the unit area of the LAGB. For a tilt LAGB one obtains

$$\rho_{tilt}^{GB} = \frac{Db - bb}{Db} + \frac{bb}{Db}\rho^D \qquad \text{B.2}$$

where $\rho^D$ is the relative atomic density at the core of an edge dislocation. Inserting Eq. B.1 in Eq. B.2

$$\rho_{tilt}^{GB} = 1 - (1 - \rho^D)\sin\theta. \qquad \text{B.3}$$

Here it is assumed that the volume between the two dislocations has the same density as bulk ($\rho^B = 1$). For an edge dislocation depicted in Fig. 8 (a), one can approximate $\rho^D \approx \frac{3}{4}\left(1 + \frac{b}{4a-b}\right)$ where $a$ is the interatomic distance normal to the edge dislocation line –Here we neglect the effect of elastic energy of the dislocation on its core density. Neglecting the second term inside the parenthesis then we have $\rho^D \approx 0.75$ that gives the atomic density of a tilt GB as

$$\rho_{tilt}^{GB} \approx 1 - \frac{\sin\theta}{4} \qquad \text{B.4}$$

In this relation, as $\theta$ approaches zero, $\rho^{GB}$ approaches $\rho^B = 1$ and the GB energy (according to Eq. (7)) decreases towards zero. Inserting Eq. B.4 into Eq. (13), we obtain

$$\gamma_{tilt} = \alpha_0 \frac{(\sin\theta)^2}{16} \qquad \text{B.5}$$

that is the tilt GB energy as a function of the misorientation angle. Equations (13), B.4 and B.5 give the relationship between the GB density, energy and misorientation angle based on the current simple analysis. From Eqs. (27) and B.4, $\Delta n_{tilt} \approx \frac{\sin\theta}{4}$ and thus the corresponding average GB excess free volume will be proportional to the misorientation angle as $\Delta V \propto \sin\theta$. Figure 8 (b) and (c) show the variation of GB density and energy as a function of the misorientation angle according to this analysis.

Although Eqs. B.4 and B.5 are based on a very simplified averaging method that does not account for the elastic energy of the dislocations, the current analysis is solid in establishing the trend in GB densities as a function of its misorientation angle, with the exception of special GBs. It is useful to show that the GB density converges to the expected GB energies in the limiting cases, consistent with the previous derivations of GB energy based on the excess free volume concept [78]. In order to connect the GB segregation isotherm with its initial misorientation angle, we can approximate by replacing the atomic density parameter using B.4. From Eqs. B.4 and (19)

$$G_{SS}^{GB} = X_A^{GB}\left(E_A^B\left(1-\frac{\sin\theta}{4}\right)^2 + (K_A^B + pV_A^B - TS_A^B)\left(1-\frac{\sin\theta}{4}\right)\right)$$
$$+ X_B^{GB}\left(E_B^B\left(1-\frac{\sin\theta}{4}\right)^2 + (K_B^B + pV_B^B - TS_B^B)\left(1-\frac{\sin\theta}{4}\right)\right) \quad \text{B.6}$$
$$+ \left(1-\frac{\sin\theta}{4}\right)^2 \Omega\, X_A^{GB} X_B^{GB} - T\Delta S_{mix}^B$$

Thus, using Eqs. B.6 and (22) the GB segregation isotherm follows

$$\frac{X_B^{GB}}{1-X_B^{GB}} = \frac{X_B^B}{1-X_B^B} \cdot \exp\left(-\frac{[\Delta E^B + \Omega]\left(\left(1-\frac{\sin\theta}{4}\right)^2 - 1\right)(\Delta K^B + p\Delta V^B - T\Delta S^B)\left(\left(1-\frac{\sin\theta}{4}\right) - 1\right) + 2\Omega\left[X_B^B - \left(1-\frac{\sin\theta}{4}\right)^2 X_B^{GB}\right]}{RT}\right) \quad \text{B.7}$$

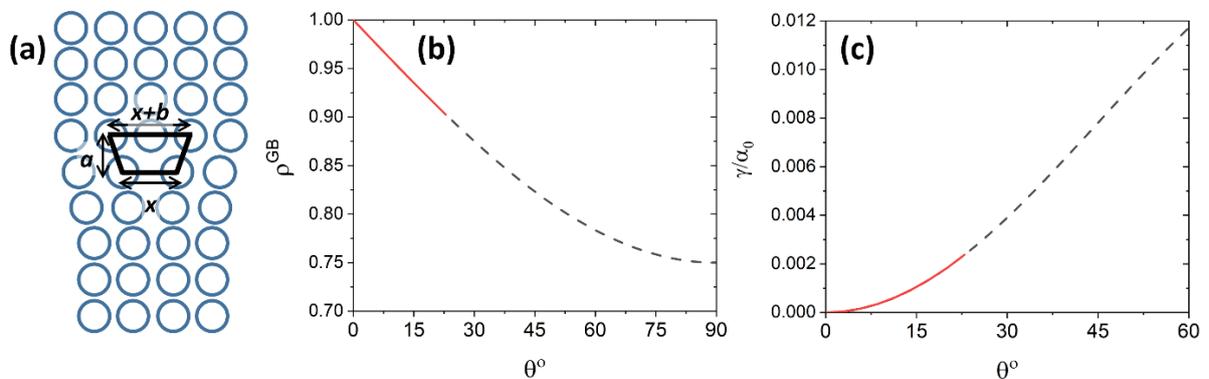

Figure 8: (a) The core of an edge dislocation has a lower density than that of the bulk that is captured by a simple geometrical analysis. GB density (b) and energy (c) as a function of the misorientation angle, with the exception of special GBs, are plotted (Eqs. B.4 and B.5).

## Acknowledgement

This work is being performed under the project DA 1655/2-1 within the Heisenberg program by the Deutsche Forschungsgemeinschaft (DFG). The author gratefully acknowledges financial supports within this program.

## Data Availability

The numerical data from this study are available upon reasonable request.